\documentclass[prc,twocolumn,superscriptaddress,showpacs,twoside,floatfix]
{revtex4-1}

\usepackage{amssymb,epsfig}

\begin{document}

\title{Thomas-Ehrman effect in a three-body model: $^{16}$Ne case}

\author{L.V.~Grigorenko}
\affiliation{Flerov Laboratory of Nuclear Reactions, JINR, RU-141980 Dubna,
Russia}
\affiliation{National Research Nuclear University ``MEPhI'', Kashirskoye shosse
31, RU-115409 Moscow, Russia}
\affiliation{National Research Center ``Kurchatov Institute'', Kurchatov sq.\ 1,
RU-123182 Moscow, Russia}
\author{T.A.~Golubkova}
\affiliation{Advanced Educational and Scientific Center, Moscow State
University, Kremenchugskaya 11, RU-121357 Moscow, Russia}
%

%\affiliation{Vasilchenko 7, RU-98680 Simeiz, Russia}
%
\author{M.V.~Zhukov}
\affiliation{Fundamental Physics, Chalmers University of Technology, S-41296
G\"{o}teborg, Sweden}

%\date{\today. {\tt File: c:/latex/16ne-tes/16ne-tes-4.tex }}

\begin{abstract}
The dynamic mechanism of the Thomas-Ehrman shift in three-cluster systems is 
studied  by example of $^{16}$Ne and $^{16}$C isobaric mirror partners. We 
predict configuration mixings for $0^+$ and $2^+$ states in $^{16}$Ne and 
$^{16}$C. Large isospin symmetry breaking on the level of wave function 
component weights is demonstrated for these states and discussed as three-body 
mechanism of Thomas-Ehrman shift. It is shown that the description of the 
Coulomb displacement energies requires a consistency among three parameters: the 
$^{16}$Ne decay energy $E_T$, the $^{15}$F ground state energy $E_r$, and the 
configuration mixing parameters for the $^{16}$Ne/$^{16}$C $0^+$ and $2^+$ 
states. Basing on this analysis we infer the $^{15}$F $1/2^+$ ground state 
energy to be $E_r=1.39-1.42$ MeV.
\end{abstract}


\maketitle

%===============================================================================

\section{Introduction}

%===============================================================================

This paper probes if the Thomas-Ehrman shift (TES) can be used to gain more 
insight in the low-lying $^{16}$Ne spectrum and in the properties of the even 
$s$-$d$ shell nuclei beyond the proton dripline in general.

The TES is an effect of the isobaric symmetry violation initially introduced for 
single-particle states of $sd$-shell nuclei in Refs.\ 
\cite{Thomas:1951,Ehrman:1951}. In such nuclei the $l=0$ and $l=2$ orbitals are 
quite close to degeneracy. However, these orbitals have different radial extent 
and their Coulomb displacement energies (CDE) are quite different. Therefore the 
relative positions of the $l=0$ and $l=2$ orbitals in isobaric partner states 
are strongly affected by the presence (absence) of the Coulomb interaction. 
These differences provide e.g.\ a simple way for $l$ identification. Later the 
studies of the TES effect have also been extended to nuclei with even number of 
``valence'' nucleons, having in mind e.g.\ understanding of configuration 
mixing.

There is no solid definition of TES, so, let us remind about those typically 
used in the literature. One possibility is the kind of ``theoretical'' 
definition which considers difference between experimental CDE and the one 
expected from isobaric symmetry
\begin{equation}
\Delta_J = \Delta_{\text{Coul}}(\text{calc}) -
\Delta_{\text{Coul}}({\text{pert}})\,,
\label{eq:tes-th}
\end{equation}
for the state with total spin $J$. The $\Delta_{\text{Coul}}(\text{calc})$ is 
CDE obtained by solving a Schr\"{o}dinger equation (SE) both on the 
\emph{proton} and \emph{neutron} sides of the isobar, while the 
$\Delta_{\text{Coul}}(\text{pert})$ is perturbative CDE obtained by solving SE 
on the \emph{neutron} side of the isobar and then using the obtained wave 
function (WF) to calculate CDE on the proton side perturbatively by assuming 
complete isobaric symmetry:
\[
\Delta_{\text{Coul}}({\text{pert}}) = \langle \Psi_n | V_{\text{Coul}} | \Psi_n
\rangle \, .
\]
Such a definition was used e.g.\ in Refs.\
\cite{Auerbach:2000,Grigorenko:2002,Garrido:2004}.

A phenomenological analogue of the value (\ref{eq:tes-th}) was analyzed in Ref.\
\cite{Comay:1988}. This work compared experimental masses $M_{\text{exp}}$ with
masses $M_{\text{cg}}$ provided by charge-symmetric mass relationship
\[
\Delta_J = M_{\text{exp}} - M_{\text{cg}}\,.
\]
Such an analysis relies only on the information about masses and excitation 
energies and it can be performed in unambiguous and statistically significant 
way. A systematic increase in the value of TES was demonstrated in 
\cite{Comay:1988} for systems beyond the proton dripline with increasing of 
proton (or two-proton) Q-values.

The other opportunity is to use a pure ``phenomenological'' definition which
relies only on the experimental relative shifts of the energy levels with
different $J$ values in proton-rich and neutron-rich mirror systems:
\begin{equation}
\Delta_{J_2 J_1} = [E(J_2) - E(J_1)]_{\text{prot}} - [E(J_2) -
E(J_1)]_{\text{neut}} \,.
\label{eq:tes-ph}
\end{equation}
An analysis of this definition of TES compared with the one of Eq.\ 
(\ref{eq:tes-th}) can be found in Ref.\ \cite{Auerbach:2000}.

The interpretation of TES for the systems with one valence nucleon is very 
simple as we have already mentioned. The energies of single particle orbitals 
with $l=2$ can be found around the values defined by the perturbative Coulomb 
displacement, and $\Delta_J$ for these should be relatively small. The energies 
of single particle orbitals with $l=0$ are shifted to considerably lower 
energies than perturbative values, and $\Delta_J$ for these should be large. As 
a result the distance between levels (and sometimes even the level ordering) is 
changing inducing a sizable $\Delta_{J_2 J_1}$.

The situation is more complicated for systems with two valence nucleons. A
simple estimate illustrating that is as follows. For $sd$-shell WFs of the $0^+$ 
and $2^+$ states can schematically be approximated as
\begin{eqnarray}
\Psi_0  & = & \alpha_0 [s^2]_0 + \beta_0 [d^2]_0 \, \nonumber ,\\
\Psi_2  & = & \alpha_2 [sd]_2 + \beta_2 [d^2]_2 \, ,
\label{eq:sd-wf}
\end{eqnarray}
If we think in terms of the independent particle model and that a typical value 
of TES associated with an $s$-wave nucleon is $\Delta$, and with $d$-wave 
nucleon zero, then e.g.\ for $\alpha_0=1$ and $\alpha_2=0$ we can expect 
$\Delta_{20}=2 \Delta$ while for $\alpha_0=0$ and $\alpha_2=1$ we can expect
$\Delta_{20}=-\Delta$. Thus, in principle, the important structure information
is encoded in the $\Delta_{J_2 J_1}$ value, but it still can not be extracted
without a considerable theoretical work, in a contrast with the one valence 
nucleon case.

The existence of specific realization of TES, characterized as a ``three-body
mechanism'' of TES, was demonstrated in Ref.\ \cite{Grigorenko:2002}. The
$^{12}$O, $^{16}$Ne, and their isobaric mirror partners $^{12}$Be, $^{16}$C were
considered in a three-body core+$N$+$N$ model. It was shown that in such
systems not only conventional (let's call it ``static'') TES, connected with
different radial extent of the $[s^2]$ and $[d^2]$ configurations exists. Also
there arises a specific TES of three-body nature (a ``dynamic'' TES) for which 
the relative weights of $[s^2]$ and $[d^2]$ configurations appeared to be
strongly different in the neutron-rich and proton-rich mirror partners. A strong
increase (tens of percent) was predicted for the weight of the $[s^2]$
configuration on the proton side of the isobar caused by the presence of the 
core-$p$ Coulomb interaction.

Recently $^{16}$Ne was studied in three experiments using neutron
knockout from a $^{17}$Ne beam \cite{Mukha:2008,Wamers:2014,Brown:2014}, and 
providing data with better statistics and quality than in the previous works. 
This inspired us to revisit the issue and consider the TES effect in 
$^{16}$Ne/$^{16}$C also in the broader context including the first excited $2^+$ 
states. We demonstrate in this work that TES can be used as a very precise tool 
to check consistency of three-body core+$N$+$N$ and two-body core+$N$ state 
properties.

%===============================================================================

\section{Theoretical model}

%===============================================================================

The theoretical model of this work is the same as was previously used for the 
discrete spectrum \cite{Danilin:1991} and continuum \cite{Pfutzner:2012} studies 
in a three-body approach. It was applied to $^{16}$Ne and its isobaric mirror
partner $^{16}$C in Refs.\ \cite{Grigorenko:2002,Brown:2014}. Here we describe
the model mainly to clarify details connected with our accurate TES treatment.

For studies of $^{16}$C, the discrete spectrum states are solutions of  a 
homogeneous three-body Schr\"odinger equation
\begin{equation}
(\hat{H}_3 - E_T)\Psi_3(\rho,\Omega_5) = 0\,,
\label{eq:sch-bs}
\end{equation}
where the energy $E_T$ is calculated with respect to the core+$N$+$N$ threshold.
For studies of the $^{16}$Ne continuum spectrum an inhomogeneous three-body
Schr\"odinger equation
\begin{equation}
(\hat{H}_3 - E_T)\Psi_3^{(+)}(\rho,\Omega_5) =
\Phi^{(J)}_{\mathbf{q}}(\rho,\Omega_5)
\label{eq:sch-cs}
\end{equation}
is solved for each $J$ different energies $E_T$ searching for the resonance peak
position. The source function $\Phi^{(0)}_{\mathbf{q}}$ for the $^{16}$Ne g.s.\
was approximated assuming a sudden removal of a neutron from the $^{15}$O core
of $^{17}\text{Ne}$,
\begin{equation}
\Phi^{(0)}_{\mathbf{q}} = v_0 \int d^3 r_n \, e^{i\mathbf{q r}_n} \langle
\Psi_{^{14}\text{\scriptsize
O}} | \Psi_{^{17}\text{\scriptsize Ne}} \rangle \, ,
\label{eq:sour-0}
\end{equation}
where $\mathbf{r}_n $ is the radius vector of the removed neutron. The
$^{17}\text{Ne}$ g.s.\ WF $\Psi_{^{17}\text{\scriptsize Ne}}$ was obtained in
\cite{Grigorenko:2003} in a three-body $^{15}$O+$p$+$p$ model  and different
aspects of nuclear dynamics for this system were investigated in
\cite{Grigorenko:2005}. The WF of the removed neutron was constructed in the
$^{14}$O+$n$ approximation for $^{15}$O in such a way that the neutron
separation energy and experimental matter radius of $^{15}$O are reproduced, see
Ref.\ \cite{Sharov:2014} for details of the whole procedure.
For the $2^+$ excitations of $^{16}$Ne we do not have
some simple dynamically motivated model and $\Phi^{(2)}_{\mathbf{q}}$ was
provided by additionally acting on the valence protons of the $^{17}$Ne g.s.\ WF
by the quadrupole
operator:
\begin{equation}
\Phi^{(2)}_{\mathbf{q}} = v_2 \int d^3 r_n e^{i\mathbf{q r}_n} \langle
\Psi_{^{14}\text{\scriptsize O}} |\sum_{i=1,2} r^2_i \, Y_{2m_i}(\hat{r}_i)| 
\Psi_{^{17}\text{\scriptsize Ne}}
\rangle \, .
\label{eq:sour-2}
\end{equation}
The sudden removal approximation is not intended for absolute cross section
calculations, therefore the ``source strength'' coefficients $v_J$ are arbitrary 
values providing the source functions the dimension of energy.

It should be noted that the approaches to discrete spectrum and continuum states 
is explicitly different in Eqs.\ (\ref{eq:sch-bs}) and (\ref{eq:sch-cs}). There
are two things to note about that. (i) The formulation provided by Eq.\
(\ref{eq:sch-cs}) is a simplistic, but reasonable approximation
for the neutron knockout reaction mechanism used to populate $^{16}$Ne states in
the recent experimental studies \cite{Mukha:2008,Wamers:2014,Brown:2014}.
Therefore, the differences in the approaches (\ref{eq:sch-bs}) and 
(\ref{eq:sch-cs}) is physically motivated. (ii) The practical difference between 
results provided by Eqs.\ (\ref{eq:sch-bs}) and (\ref{eq:sch-cs}) vanishes in 
the limit widths tending to zero for continuum states. For $0^+$ and $2^+$ 
states of $^{16}$Ne considered in this work (which are very narrow), the effect 
of particular choice of the sources
$\Phi^{(J)}_{\mathbf{q}}$ on the energies of the calculated resonances is less 
than some units of keV, which is much less than the other effects considered 
here.

The three-body WF $\Psi_3$ depends on a set of hyperspherical variables:
the hyperradius $\rho $ and the five-dimensional hyperangle $\Omega_5$. The
hyperspherical decomposition of the discrete spectrum  WF is
\begin{equation}
\Psi_3(\rho,\Omega_5) = \rho^{-5/2} \sum_{K\gamma} \chi_{K\gamma}(\rho)
\mathcal{J}_{K\gamma}(\Omega_5)\,.
\label{eq:psi-bs}
\end{equation}
The value $K$ is the hypermoment (the principal quantum number of the 
hyperspherical method) while the ``multiindex'' $\gamma=\{L,S,l_x,l_y\}$ stands 
for the complete set of quantum numbers for the specific three-body WF 
component: total orbital momentum $L$, total spin $S$, and orbital angular 
momenta $l_x$ and $l_y$ for the Jacobi subsystems. The boundary conditions for 
the discrete spectrum partial hyperspherical functions $\chi_{K\gamma}$ are 
expressed in terms of Bessel functions $K$,
\begin{eqnarray}
\chi_{K\gamma}(\rho) & \stackrel{\rho \rightarrow \infty}{\sim} & \sqrt{
2\varkappa \rho / \pi} \, K_{K+2} (\varkappa \rho) \nonumber \\
& \sim & \exp[-\varkappa \rho] \left( 1 + \textstyle \frac{4(K+2)^2-1}{8
\varkappa \rho} + \ldots \right) \,,
\label{eq:psi-bs-ass}
\end{eqnarray}
where $\varkappa=\sqrt{2ME_T}$, and $M$ is the average mass of the
nucleon in the considered system. At the large (tens of Fermi) distances this is
essentially an exponential decrease.

The decomposition of the continuum WF
$\Psi_3^{(+)}$ is analogous to that of Eq.\ (\ref{eq:psi-bs}). The boundary
conditions for the partial continuum functions $\chi^{(+)}_{K\gamma}$ at
extreme remote asymptotic hyperradius (where the long-range Coulomb terms vanish 
because of some form of physical screening) should be provided by diverging 
waves:
\begin{equation}
\chi^{(+)}_{K\gamma}(\rho) \stackrel{\rho \rightarrow \infty}{\sim}
\exp[i\varkappa \rho] \,.
\label{eq:psi-cs-ass}
\end{equation}
However, for realistic distances of actual calculations  (e.g.\ $\rho \sim 1000$
fm) we use complicated approximate boundary conditions for the three-body
Coulomb problem obtained by diagonalization of the Coulomb potential terms in
the hyperspherical representation on the finite hyperspherical basis
\cite{Pfutzner:2012}.

The three-body Hamiltonian consists of the kinetic term, three pairwise
interactions, and the phenomenological potential $V_3$ depending only on the
hyperradius $\rho$:
\begin{equation}
\hat{H}_3 = \hat{T} + V_{\text{core-}N_1} + V_{\text{core-}N_2} +
V_{N_1\text{-}N_2} + V_3(\rho)\,.
\label{eq:ham}
\end{equation}
The $V_3$ term \emph{technically} aims at fine correction of the state energies
when we need to adjust them exactly to experimental values, and
\emph{physically} accounts for many-body effects which are beyond the
three-body dynamics \cite{Pfutzner:2012}. In this work we use for this term
the Woods-Saxon formfactor
\begin{equation}
V_3(\rho) = V^{(J)}_3/(1+\exp[(\rho-\rho_0)/a_{\rho}])\,,
\label{eq:v3}
\end{equation}
with the radius $\rho_0=6$ fm and diffuseness $a_{\rho}=0.6$ fm. The $V_3$
potential depth parameter $ V^{(J)}_3$ is adjusted individually for each total 
spin $J$ (see Table \ref{tab:poten}, for example).

For the $^{14}Z$+$N$ channel ($Z=9,6$) we use the potentials very similar to 
those from Refs.\  \cite{Grigorenko:2002,Brown:2014}, but with minor variations 
connected with the procedure of the TES treatment discussed in the next 
paragraph. These are Woods-Saxon potentials with derivative $(ls)$ term
\begin{eqnarray}
V_{\text{core-}N_i} = V^{(l)}_{c}\frac{1}{1+f(r)} + (\mathbf{l} \cdot
\mathbf{s}) \, V^{(l)}_{ls} \frac{b}{ra}\frac{f(r)}{[1+f(r)]^2} \, ,\nonumber \\
f(r) = \exp[(r-r^{(l)}_0)/a] \, ,\nonumber
\label{eq:v}
\end{eqnarray}
where $b=2.01532$ fm$^2$. The components of the potentials are $l$-dependent: 
they are adjusted individually for the quantum states with different angular 
momenta. The following parameters are used for $s$, $p$,
and $d$ orbitals: $a=0.53$ fm, $V^{(1)}_{c}=-12$ MeV, $V^{(1)}_{ls}=11$ MeV,
$r^{(1)}_0=2.89$ fm,  $V^{(2)}_{ls}=-11.12$ MeV, $r^{(2)}_0=3$ fm. The other
central component parameters for $s$- and $d$-waves are provided in Table
\ref{tab:poten}. There is also an additional repulsive component in the $s$-wave
with Woods-Saxon formfactor with repulsion 144 MeV, width 1.7 fm and
diffuseness 0.53 fm. This is required to simulate the effect of the occupied
deep $s$ orbital in the $^{14}Z$ core cluster in our three-body model.

In this work we employ for the nucleon-nucleon channel a quasirealistic
potential Ref.\ \cite{Gogny:1970} including central, spin-orbit, tensor, and
parity splitting terms.

The Coulomb potential of a homogeneously charged sphere was used in this work
with sphere radius adjusted to reproduce the specific charge radius. We also
estimated the influence of the charge distribution on the effect of using
Gaussian and Fermi type formfactors. The impact of such a modification on
energies is on the level $10-15$ keV, which is much smaller than the scale of
energy uncertainty connected with uncertainty of the charge radius itself.

In the large-basis calculations we treat part of the basis adiabatically.
The potential matrix of the large size $K_{\max}$ is reduced to the size
$K_{FR}$ by a procedure  which is called Feschbach reduction. This is 
essentially an adiabatic approximation, see, e.g., Ref.\ \cite{Grigorenko:2009c} 
for details. A reduced potential matrix of size $K_{FR}$ is used to solve the 
system of hyperspherical coupled channel equations. A $K_{\max}$ equal to 110 
and 70 is used for the $0^+$ and $2^+$ states correspondingly, while the 
$K_{FR}$ values are 24 and 18. Such basis sizes are sufficient for computational 
convergence of such complicated decay observables as widths and momentum
distributions \cite{Grigorenko:2009c,Brown:2014}. They are more than sufficient
for full convergence of the energies calculations.

%===============================================================================

\section{Computation procedure for TES}

%===============================================================================

\begin{figure}
\begin{center}
\includegraphics[width=0.47\textwidth]{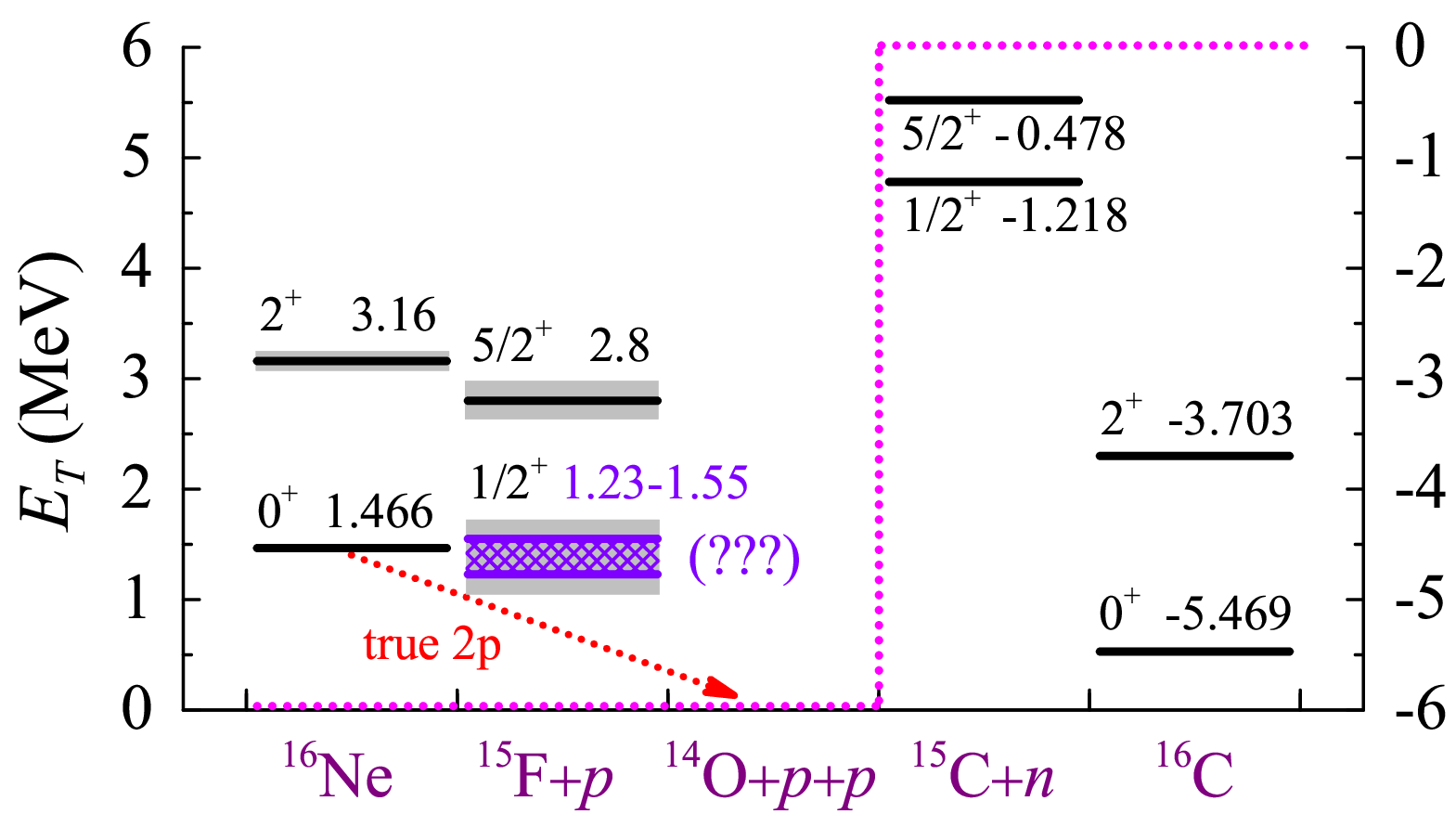}
\end{center}
\caption{Energy level schemes for $^{16}$Ne (from \cite{Brown:2014}), $^{15}$F
(left axis) and $^{16}$C, $^{15}$C (right axis). The true $2p$ decay path for
the $^{16}$Ne g.s.\ is indicated by the red dotted arrow. Hatched area indicates
the experimental uncertainty of the $^{15}$F g.s.\ position.}
\label{fig:levels}
\end{figure}

There are two uncertain ingredients in the three-body model computation of
TES in the $^{16}$Ne/$^{16}$C mirror partner pair: (i) the charge radius
$r_{\text{ch}}$  of $^{14}$O, which is experimentally unknown and (ii) the
$1/2^+$ ground state energy $E_r$ of $^{15}$F, which strongly affects the
results of the calculations, but on which there exists an experimental
controversy, see, e.g.\ Ref.\ \cite{Fortune:2006}. The level schemes for
$^{16}$Ne/$^{16}$C and $^{15}$F/$^{15}$C isobaric mirror partner pairs are
shown in Fig.\ \ref{fig:levels}.

%-------------------------------------------------------------------------------
%-------------------------------------------------------------------------------

Preparing the potential sets we fix the positions of $1/2^+$ and $5/2^+$ states 
of $^{15}$C$=^{14}$C+$n$ to be exactly experimental values by using potentials 
of somewhat different radii $r_0$ in the corresponding partial wave. Then we 
switch to $^{15}$F getting different resonant state positions $E_r(J^{\pi})$ 
depending on the potential radius and also on $r_{\text{ch}}$ of $^{14}$O. The 
$d$-wave potential radius is thus fixed based on the well known position of the 
$5/2^+$ state in $^{15}$F. The $1/2^+$ state position in $^{15}$F is varied 
depending on specific $r_0$ and $r_{\text{ch}}$. For relatively broad states, to 
which the $^{15}$F g.s.\ belongs, there is always some uncertainty in the 
definition of the state position. In this work we always imply that $E_r$  is 
the energy at which the phase shift passes $\pi/2$.

With the obtained potential sets we run three-body model calculations for
$^{16}$C. The phenomenological three-body potential parameters $V^{(J)}_{3}$
are adjusted to provide exact experimental energies of $0^+$ and $2^+$ states
of $^{16}$C, $E_T(0^+)=-5.469$ MeV and $E_T(2^+)=-3.703$ MeV. With
$V^{(J)}_{3}$ parameters adjusted in this way we then perform the calculations
of $^{16}$Ne. The results for several potential sets (P1, P2, P3) giving
different $^{15}$F g.s.\ positions, are provided in Table \ref{tab:poten}. In
the Table we show only the calculation inputs and results for the charge radius
$r_{\text{ch}}(^{14}\text{O})=2.7$ fm since the values obtained with different
charge radii differ insignificantly.

It is seen in Table \ref{tab:poten} that for realistic potentials P1-P3
the values of $V^{(J)}_{3}$ are quite small, typically below 1 MeV. They are
also quite similar for both $0^+$ and $2^+$ states. This indicates that 
three-body picture of the inert $^{14}$O/$^{14}$C core plus
two nucleons is an adequate approximation to the structure of these low-lying
states in $^{16}$Ne/$^{16}$C.

To test the sensitivity of TES to the structure, we varied the $s/d$ ratio of 
the WF by the following procedure: We increased the $[s^2]$ content of the WF by 
multiplying the $d$-wave potential depth parameter $V^{(2)}_c$  by a factor 
smaller than unity, and vice versa to increase the $[d^2]$ content of the WF we 
multiply the $s$-wave potential depth parameter $V^{(0)}_c$ by a factor smaller 
than unity. Two limiting cases of such potential sets (P4, P5) are illustrates 
in Table \ref{tab:poten}. The P4 results for the $2^+$ state are missing there 
as it is not possible to construct a low-lying $2^+$ only on the $s$-wave 
orbitals.

%===============================================================================
\begin{table}[b]
\caption{Potential sets in the core+$N$ channel adjusted for
$r_{\text{ch}}(^{14}\text{O})=2.7$ fm. Radii are in fm, energies in MeV, and
probabilities in percent. The position $E_r$ of the two-body resonance is
defined here by the phase shift equals $\pi/2$. The energy of the first excited
state of
$^{15}$F is $E_r(5/2^+)=2.8$ MeV.}
\begin{ruledtabular}
\begin{tabular}[c]{cccccc}
  & P1 & P2 & P3 & P4 & P5  \\
\hline
$E_r(1/2^+)$    &  1.147   & 1.287    & 1.467    & 1.287    & 1.287     \\
$r^{(0)}_0$     &  3.5     &  3.1     & 2.7      & 3.1      &  3.1      \\
$V^{(0)}_c$     & -34.085  & -47.45   & -67.9    & -47.45   &  0        \\
$V^{(2)}_c$     & -49.587  & -49.587  & -49.587  & 0        & -49.587   \\
\hline
$V^{(0)}_{3}$   & 0.255    & 0.647    & 1.051    & -2.461   & -1.918    \\
$\Delta_{\text{Coul}}({\text{pert}})$
                &  7.017   & 7.130    & 7.301    &  6.517   &  7.561    \\
\multicolumn{6}{c}{$^{16}$C($0^+$) }    \\
$W(s^2)$        &  44.2    & 47.8     & 51.4     & 93.5     & 0.82      \\
$W(p^2)$        &  0.88    & 0.86     & 0.83     & 5.47     & 9.01      \\
$W(d^2)$        &  46.5    & 43.1     & 39.9     & 0.71     & 89.8      \\
\multicolumn{6}{c}{$^{16}$Ne($0^+$)  }      \\
$E_T$           &  1.136   & 1.303    & 1.514    & 0.821    & 1.972     \\
$\Gamma$ (keV)  &  0.323   & 1.09     & 3.67     & 0.003    & 0.313     \\
$W(s^2)$        &  71.7    & 70.2     & 69.0     & 95.4     & 1.16      \\
$W(p^2)$        &  5.88    & 5.84     & 5.98     & 3.67     & 9.59      \\
$W(d^2)$        &  22.2    & 23.4     & 24.6     & 0.62     & 88.8      \\
\hline
$V^{(2)}_{3}$   & 0.355    & 0.72     & 1.09     &          & -2.697    \\
$\Delta_{\text{Coul}}({\text{pert}})$
                &  6.927   & 7.062    & 7.229    &          &  7.436    \\
\multicolumn{6}{c}{$^{16}$C($2^+$)} \\
$W(p^2)$        &   4.67   & 4.79     & 4.9      &          & 2.50      \\
$W(d^2)$        &  15.9    & 14.6     & 13.3     &          & 94.8      \\
$W(sd)$         &  78.1    & 78.4     & 80.6     &          & 2.07      \\
\multicolumn{6}{c}{$^{16}$Ne($2^+$)}       \\
$E_T$           &  2.941   & 3.074    & 3.232    &          & 3.585     \\
$\Gamma$ (keV)  &  40.6   & 45.2      & 51.4     &          & 5.25      \\
$W(p^2)$        &  3.72    & 3.89     & 4.06     &          & 3.08      \\
$W(d^2)$        &  7.72    & 7.81     & 7.87     &          & 91.6      \\
$W(sd)$         &  87.2    & 86.9     & 86.7     &          & 4.43      \\
\end{tabular}
\end{ruledtabular}
\label{tab:poten}
\end{table}
%===============================================================================

%===============================================================================

\section{Calculational results}

%===============================================================================

%===============================================================================

\subsection{Charge radius dependence}

%===============================================================================

\begin{figure}
\begin{center}
\includegraphics[width=0.47\textwidth]{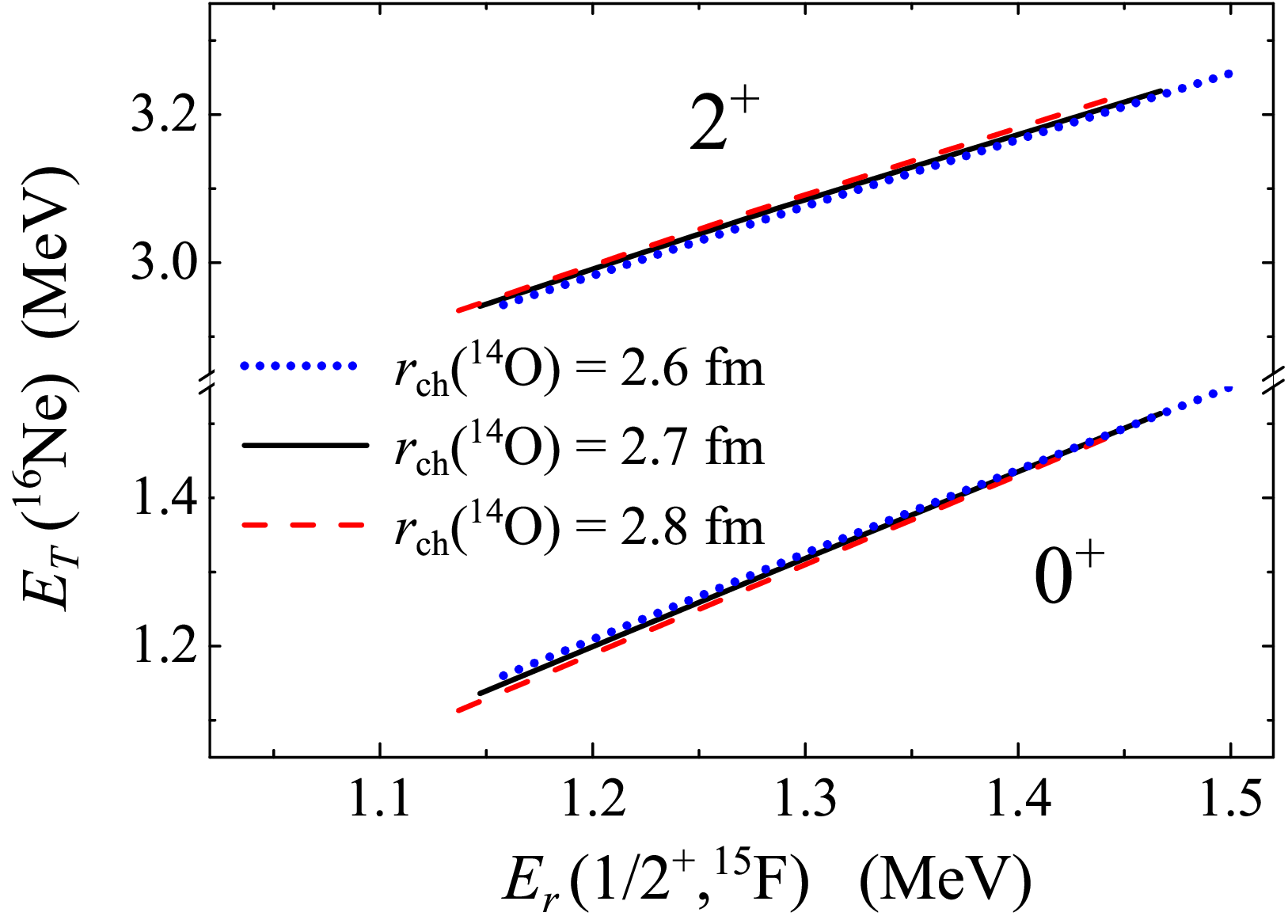}
\end{center}
\caption{Dependence of the $^{16}$Ne $0^+$  and $2^+$ state energies on the
position of the g.s.\ $1/2^+$ resonance in $^{15}$F. Calculations for potential 
sets with different charge radii of $^{14}$O are shown to be nearly 
overlapping.}
\label{fig:et-ot-er-rch}
\end{figure}

Fig.\ \ref{fig:et-ot-er-rch} shows that calculations with potentials
providing different positions of the $^{15}$F g.s.\ produce very different
positions of both $0^+$ and $2^+$ states of $^{16}$Ne. For variation of
$E_r(1/2^+)$ within the range $1.23-1.56$ MeV, ``allowed'' by uncertainty in
existing experimental data, the three-body resonance energies $E_T$ variation in
$^{16}$Ne is $250-300$ keV. In contrast, the predicted curves for different
$^{14}$O charge radii practically overlap (the deviations are less than 15
keV). Thus the influence of the specific value of the unknown charge radius of
$^{14}$O on the physically motivated calculation results (those with fixed
$^{15}$F g.s.\ position) is practically negligible and will not significantly
affect the conclusions of this work concerning TES.

%-------------------------------------------------------------------------------
%-------------------------------------------------------------------------------

%===============================================================================

\subsection{Three-body mechanism of TES}

%===============================================================================

Paper \cite{Grigorenko:2002} demonstrated that there are two major sources of 
TES in the $0^+$ ground states of even $sd$-shell nuclei present in the 
three-body core+$N$+$N$ approximation: (i) Conventional ``static'' TES connected 
with larger spatial extent of the $s$-wave orbitals compared to the $d$-wave 
orbitals, and (ii) ``dynamic'' three-body mechanism of TES leading to a relative 
increase of the $[s^2]_0$ configuration weight compared to that of the $[d^2]_0$ 
configuration.

\begin{figure}
\begin{center}
\includegraphics[width=0.47\textwidth]{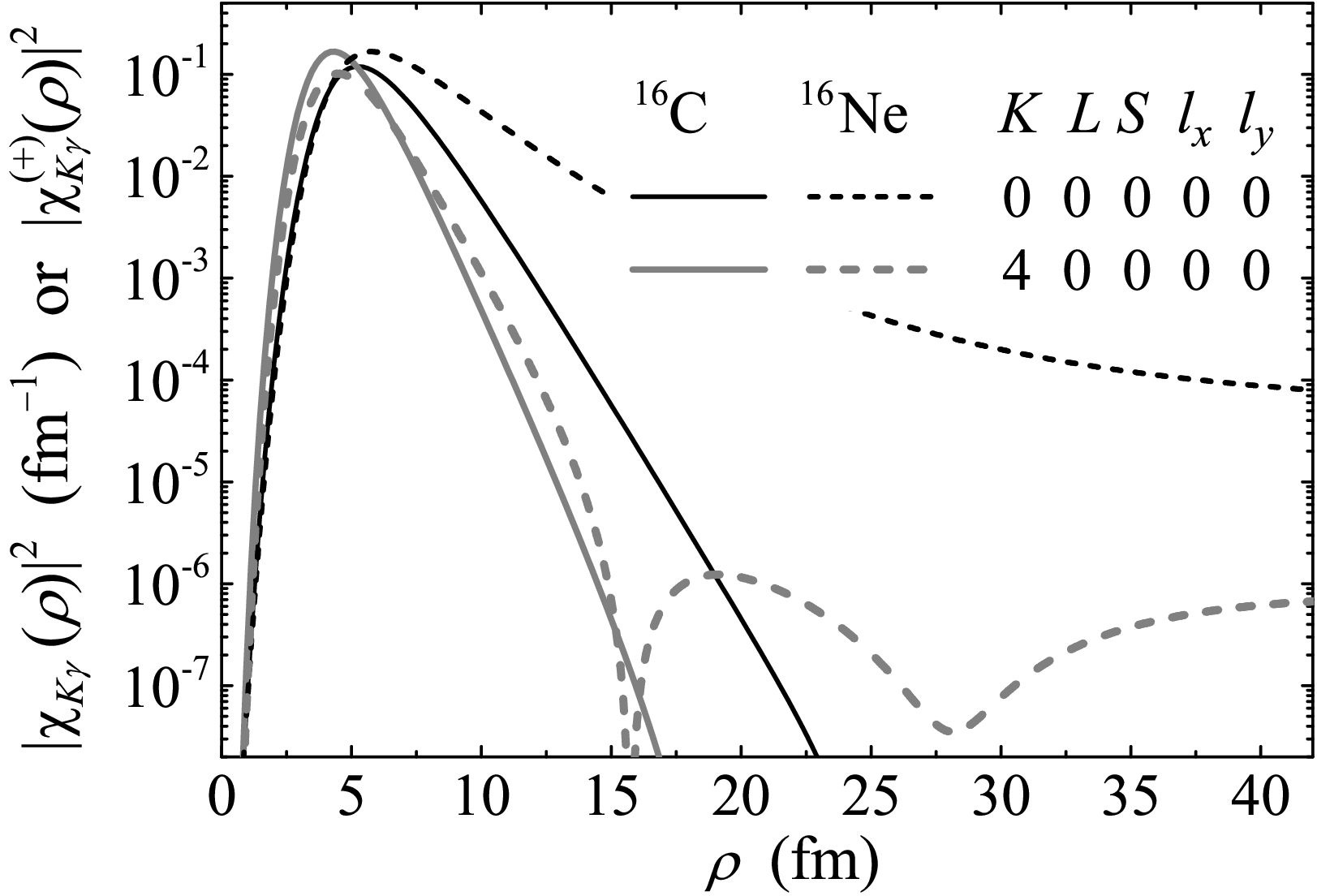}
\end{center}
\caption{Radial density dependence of two major component of three-body
WF's for the ground $0^+$ states of $^{16}$Ne and $^{16}$C. Potential set P2.
The $^{16}$C WF is normalized to unity, the $^{16}$Ne WF is normalized
arbitrarily for comfort of the visual perception.}
\label{fig:wfs}
\end{figure}

Both effects are illustrated by Fig.\ \ref{fig:wfs}, which shows the radial 
density dependence for two dominant components of $^{16}$Ne and $^{16}$C g.s.\ 
WFs. The $K=0$ component weight is very close to that of the $[s^2]_0$ 
configuration and the selected $K=4$ component corresponds well to the 
$[d^2]_0$. The densities of $^{16}$C WF components at large hyperradii 
demonstrate behavior which is close to exponential decrease. The densities of 
$^{16}$Ne WF components tend to become constant at large hyperradii, which 
corresponds to the $\sim \exp[i \varkappa \rho ]$ asymptotic of the  WF 
$\Psi_3^{(+)}$.  The radial extent of the $[d^2]_0$ component in $^{16}$Ne is a 
bit larger, but close to that in $^{16}$C. In contrast, the $[s^2]_0$ component 
is drastically broader in $^{16}$Ne. Also the weight of the $[s^2]_0$ component 
in $^{16}$Ne is evidently larger than in $^{16}$C, while the weight of the 
$[d^2]_0$ component is smaller.

The relative scale of ``static'' and ``dynamic'' TES effects can be understood
from Table \ref{tab:poten}. The calculations with potential sets P4 and P5
provide limiting cases (practically pure $s$-wave or pure $d$-wave)  of
$^{16}$Ne/$^{16}$C  structure which are very ``robust'' and are not altered by
the Coulomb interaction. Thus the TES value $\Delta_0$ for the $0^+$ state
associated solely with radial size increase of orbitals from
$^{16}$C to $^{16}$Ne is $\sim 230$ keV for pure $[s^2]$ and $\sim 120 $ keV for
pure $[d^2]$ configurations. In contrast, the predicted TES for the realistic
structure of $^{16}$Ne/$^{16}$C also includes the dynamic effect of structure
modification and thus varies between $\sim 300-400$ keV. Therefore we can 
estimate the scale of the ``dynamic'' contribution as $45-60 \%$ of the whole 
TES.

%-------------------------------------------------------------------------------
%-------------------------------------------------------------------------------

%===============================================================================

\subsection{The $^{15}$F ground state issue}

%===============================================================================

\begin{figure}
\begin{center}
\includegraphics[width=0.47\textwidth]{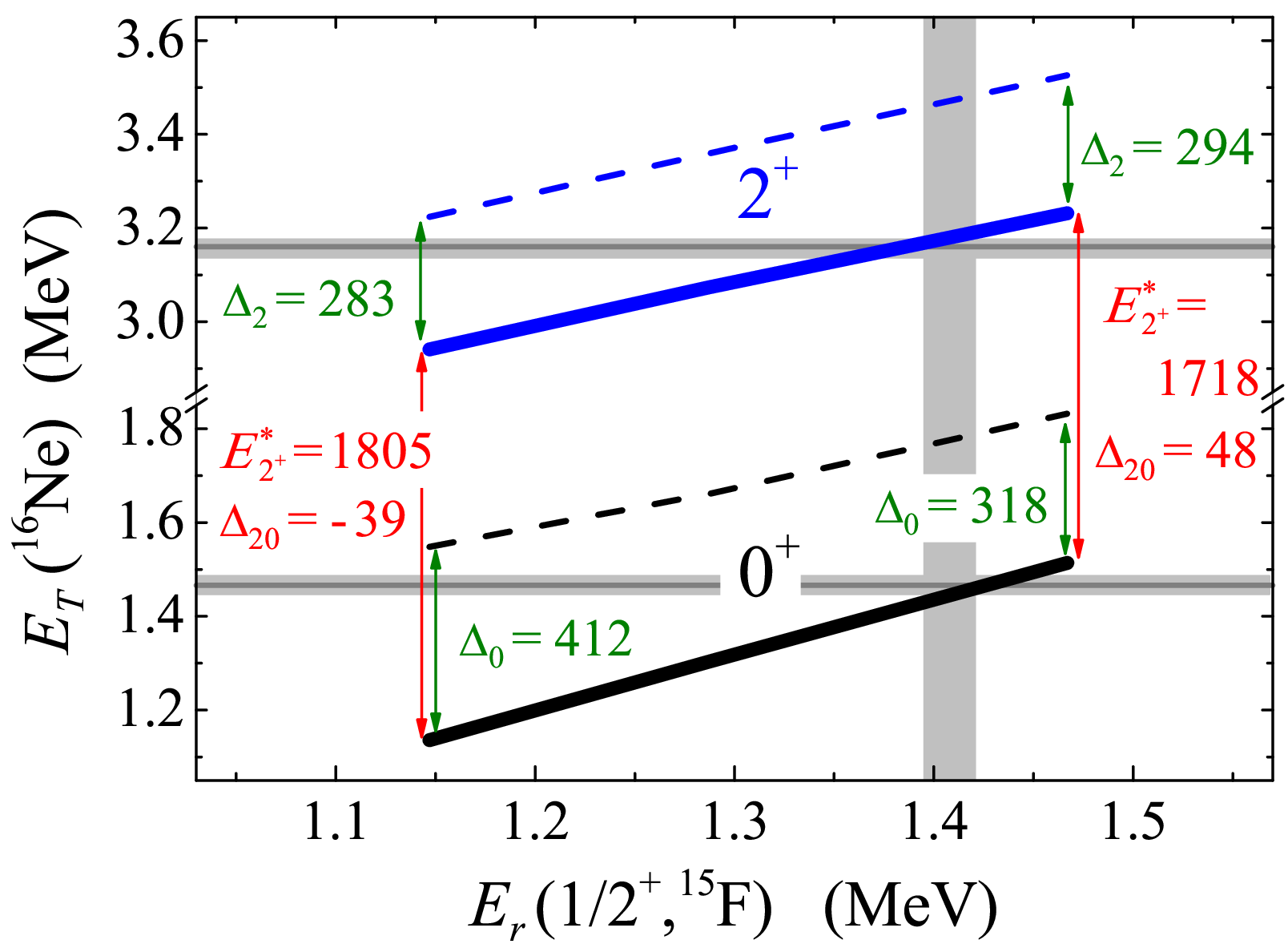}
\end{center}
\caption{Dependence of the $0^+$  and $2^+$ state energies in $^{16}$Ne on the
position of the g.s.\ $1/2^+$ resonance in $^{15}$F for dynamical (solid lines)
and perturbative (dashed lines) calculations. The thickness of theoretical
curves corresponds to $\sim 15 $ keV uncertainty connected to the choice of the
charge distribution formfactor. Horizontal gray lines and shaded areas show the
experimental $E_T$ values for $0^+$ and $2^+$ states from Ref.\
\cite{Brown:2014} with their uncertainties. The vertical shaded area indicates
the range of $E_r$ where a consistency between the theory and the experiment is
achieved. The $2^+$ state excitation energies $E^*_{2^+}$ and the TES values
corresponding to different definitions provided by Eqs.\ (\ref{eq:tes-th}) and
(\ref{eq:tes-ph}) are indicated in the plot.}
\label{fig:tes-ot-er}
\end{figure}

The calculated positions of the $0^+$ and $2^+$ states in $^{16}$Ne as function
of the $^{15}$F g.s.\ energy are shown in Fig.\ \ref{fig:tes-ot-er}. This
figure also compares dynamical and perturbative results.

``Theoretical'' TES values $\Delta_J$ are always large in our calculations both
for $0^+$ state (varies between $\sim 300$ and $\sim 400$ keV) and  for $2^+$
state (stable at about $ \sim 300$ keV) in $^{16}$Ne. In contrast, the
``phenomenological'' TES value $\Delta_{J_1J_2}$ is small and even changes sign,
being sensitive to the particular value of $E_r$ in $^{15}$F. The latter result
can probably be qualitatively understood as follows: If we look in Table
\ref{tab:poten} the weight $W(s^2)$ varies $44-51 \%$ for $0^+$ in $^{16}$C,
while the  $W(sd)$ is more stable around $79 \%$ for $2^+$. Thus, there are two
$s$-wave nucleons in $0^+$, which are subject to strong TES, but the weight of
this configuration is mainly under $50\%$. There is only one $s$-wave nucleon in
the $[sd]$ configuration of $2^+$, however, the weight of this configuration is
about two times larger in $2^+$ than $W(s^2)$ in $0^+$. So, for such structures 
of $0^+$ and $2^+$ the TES modifications of CDE are about to be equal.

Fig.\ \ref{fig:tes-ot-er} shows that simultaneous consistent theoretical 
description for both $0^+$ and $2^+$ resonances \cite{Brown:2014} is achieved at 
$E_r = 1.39-1.42$ MeV. For theoretical calculations in \cite{Brown:2014} we used 
a potential giving $E_r = 1.45$ MeV. We see now that this value is a bit too 
large considering the TES results of this work. However, slight modification of 
this parameter on such a level does not lead to any modification of conclusions 
of Ref.\ \cite{Brown:2014} related to basic theory.

Considering the high uncertainty of the current experimental situation for  
$^{15}$F g.s.\ (see, e.g., \cite{Fortune:2006}), the consistency check found in 
this work may provide now the most reliable information on the $^{15}$F g.s.\ 
energy.

%-------------------------------------------------------------------------------
%-------------------------------------------------------------------------------

\begin{figure}
\begin{center}
\includegraphics[width=0.256\textwidth]{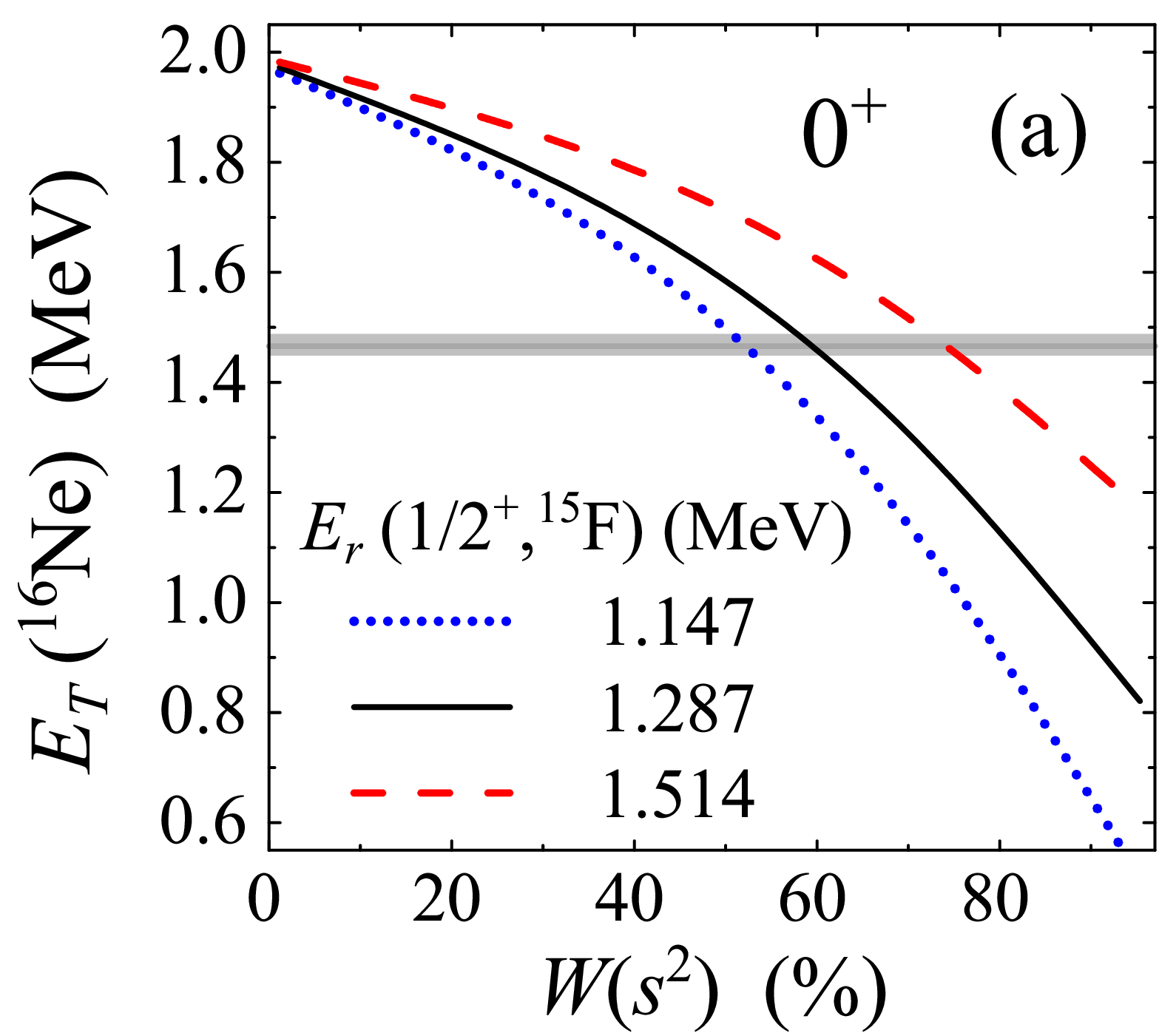}
\includegraphics[width=0.22\textwidth]{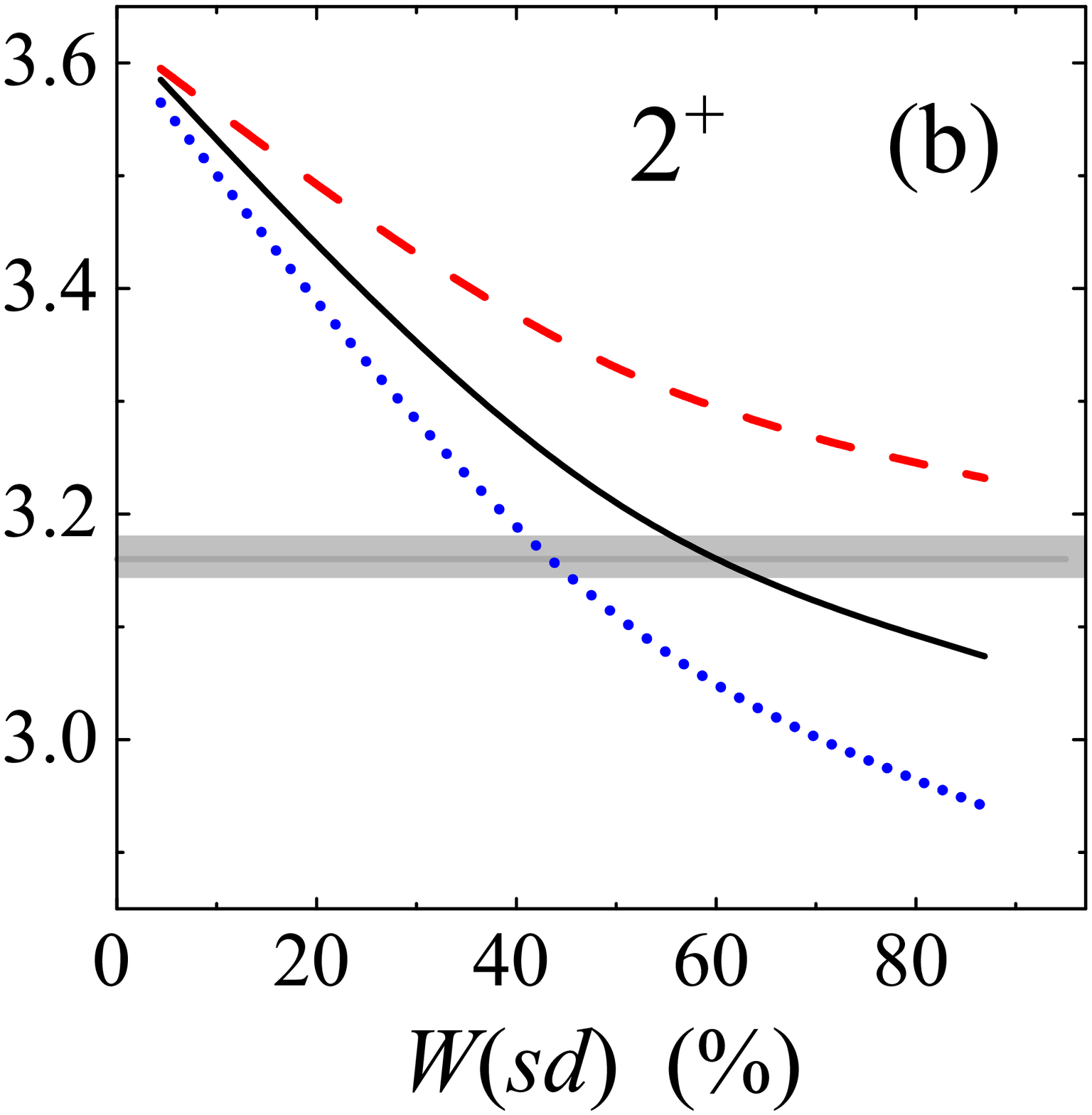}
\end{center}
\caption{Dependence of the $^{16}$Ne $0^+$ state energy on (a) the weight of the
$[s^2]$ WF component, and of the $2^+$ state energy on (b) the weight of the
$[sd]$ WF component. Horizontal gray lines and shaded areas show the
experimental values for $0^+$ and $2^+$ states from Ref.\ \cite{Brown:2014} with
corresponding uncertainty.}
\label{fig:dep-et-ot-ws2}
\end{figure}

The positions of the $0^+$ and $2^+$ states in $^{16}$Ne as functions of the 
internal structure are shown in Fig.\ \ref{fig:dep-et-ot-ws2}. Such curves allow 
to fix the configuration mixing values in the case when the decay energy $E_T$ 
is known. However, in our calculations we demonstrate that this can be made 
differently, depending on the particular $1/2^+$ resonance energy in $^{15}$F. 
Figure \ref{fig:dep-et-ot-ws2} shows that for the $0^+$ state the consistency 
with the experiment \cite{Brown:2014} is achieved for broad range (from $50\%$ 
to $75\%$) of possible configuration mixing $W(s^2)$ values, depending on the 
specific g.s.\ energy $E_r$ of $^{15}$F. Consistency with experiment 
\cite{Brown:2014} for the $2^+$ state can be achieved from $40\%$ to $100\%$ of 
possible $W(sd)$ values. However, for the $2^+$ state the situation is more 
restrictive as for $E_r \gtrsim 1.45$ MeV a consistency of TES with the 
experimental energy can not be achieved at all.

\begin{figure}
\begin{center}
\includegraphics[width=0.2525\textwidth]{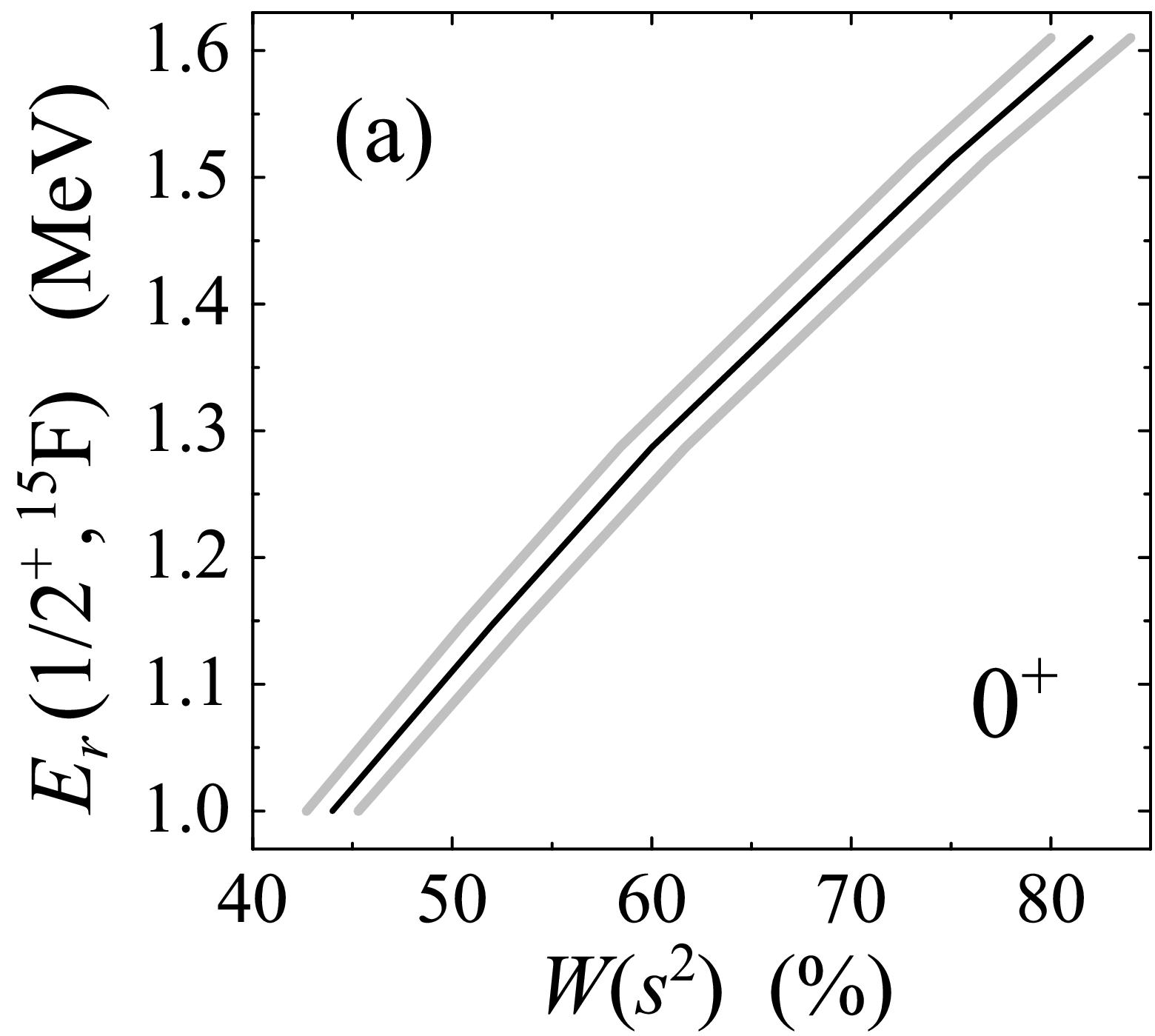}
\includegraphics[width=0.222\textwidth]{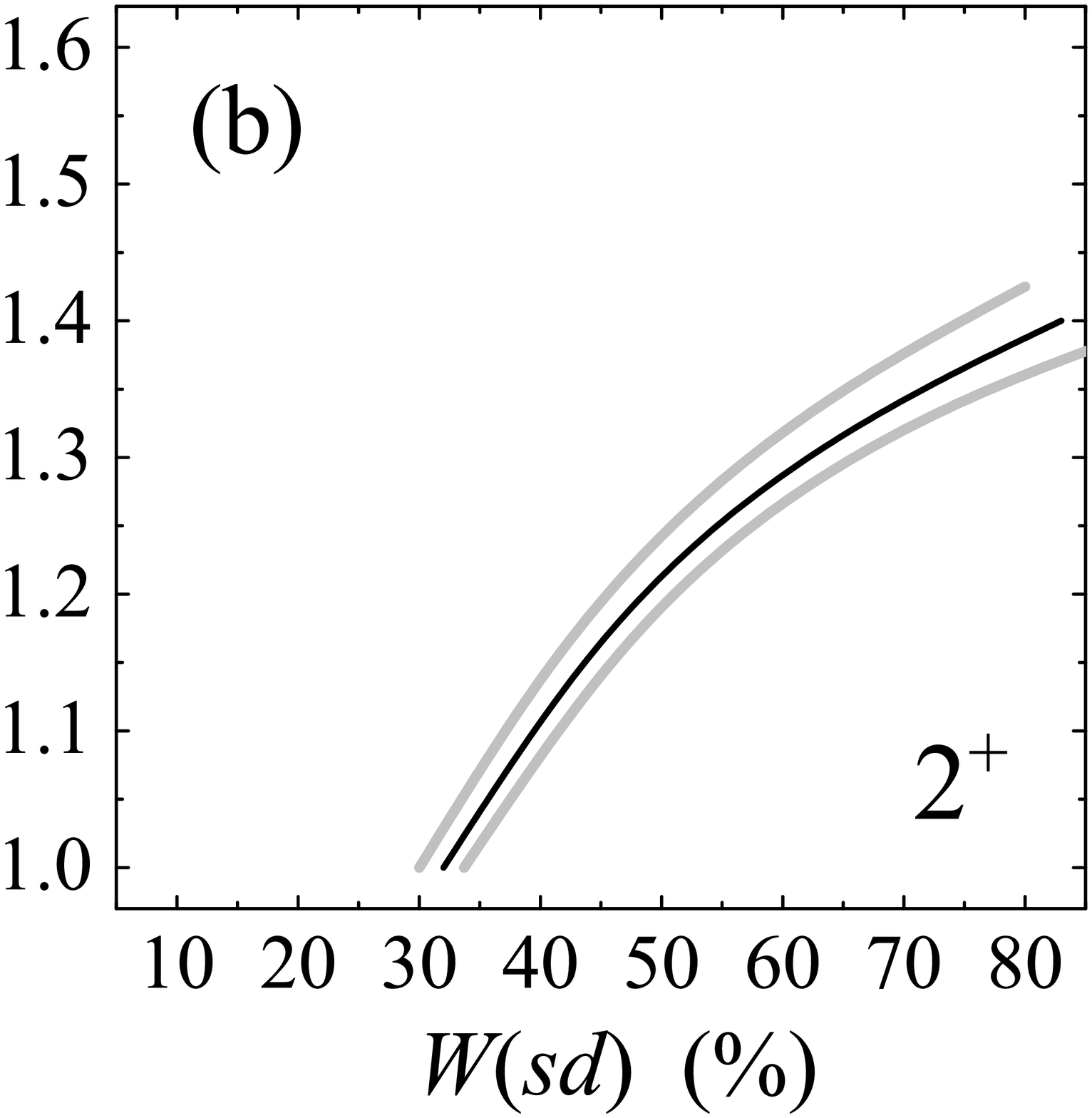}
\end{center}
\caption{Two-dimensional $\{W, E_r \}$ plots for $^{16}$Ne. (a) For the $[s^2]$ 
for $0^+$ and (b) $[sd]$ for $2^+$ WF's component weights $W(s^2)$ and
$W(sd)$ and the energy of the $^{15}$F g.s.\ under the restriction to reproduce
exactly the experimental energies of the $^{16}$Ne g.s.\ $E_T=1.466(20)$ MeV and
$2^+$ state $E_T=3.16(2)$ MeV \cite{Brown:2014}. Gray curves correspond to the
limits defined by the experimental uncertainty of the above energies.}
\label{fig:consist}
\end{figure}

If we fix the three-body state energies $E_T$ to be exactly experimental
\cite{Brown:2014} we can get the region on the planes $\{ W(s^2),E_r \}$ or $\{
W(sd),E_r \}$  where mutually consistent values of  $W$ and $E_r$ are located,
see Fig.\ \ref{fig:consist}. The meaning of these plots is that precisely fixing
the g.s.\ properties of $^{15}$F fixes the structure of the valence
configurations both for $0^+$ and $2^+$ states of $^{16}$Ne simultaneously. We
note again that while for the $0^+$ state a consistency in principle can be
achieved for a broad range of $E_r$ which is much broader than existing
experimental uncertainty, for the $2^+$ state this kind of plot becomes quite
restrictive, limiting a possible $E_r$ value to be less than $1.43-1.45$ MeV.
Thus we see that simultaneous studies of the TES for $0^+$ and $2^+$ states
imposes stringent limitations on possible properties of the core+$N$+$N$ systems
and its core+$N$ subsystems.

%-------------------------------------------------------------------------------
%-------------------------------------------------------------------------------

The value $E_r = 1.39-1.42$ MeV deduced in this work for the $^{15}$F g.s.\
appears very precise. Still there are two inherent uncertainties: (i) The 
experimental uncertainty of
$E_T$ and (ii) the theoretical uncertainties of the three-body $^{14}$O+$p$+$p$
model for $^{16}$Ne. Theoretical uncertainties which go beyond the
three-body formulation for $^{16}$Ne should further increase this
uncertainty. Thus, admixture of configurations like
$^{14}$O$^{*}$+$p$+$p$ should lead to CDE increase (these configurations should
be more compact than the main one) and thus shifts the consistency range to
somewhat lower $E_r$ values. Considering the good description of energies of
$^{16}$Ne/$^{16}$C in our model we do not expect large admixture of such
configurations and making simple estimates we can suggest a $15-30$ keV decrease
of the lower boundary of $E_r$ for $10-20 \%$ admixture of configurations with
excited $2^+$ state of the $^{14}$O core. This would extend the boundaries for
the ``TES based'' $^{15}$F g.s.\ position to $E_r = 1.36-1.42$ MeV.

%===============================================================================

\subsection{Structure of $^{16}$Ne $0^+$ and $2^+$ states}

%===============================================================================

The variation of the $^{15}$F g.s.\ energy in the range allowed by the current
experimental uncertainty produces some variations in the structure of
$^{16}$Ne and $^{16}$C states calculated in the three-body model. These
were found to be the largest for the $^{16}$C ground state, where
$W(s^2)$ varies on the level $6-8 \%$. For its isobaric mirror
partner, this uncertainty is significantly reduced. In $^{16}$Ne the typical
level of structure variation is $2-3 \%$ and can be regarded as insignificant.
We can guess that for $^{16}$Ne the peripheral dynamics, associated with the
long-range Coulomb interaction rather than with the short-range nuclear
dynamics, is more important and this leads to the relative stabilization of the
calculational results for the $0^+$ state in $^{16}$Ne.

%-------------------------------------------------------------------------------
%-------------------------------------------------------------------------------

For the $2^+$ state the predictions are much more stable than for the $0^+$
state and also follow the trend discussed above: There is $\sim 2.5 \%$
variation of $W(sd)$ in the $^{16}$C WF and just $\sim 0.5 \%$ variation in the
$^{16}$Ne WF. Both values can be regarded as very small and the predicted
structure as stable.

Quite a paradoxical output of our studies is that the presence of the Coulomb
interaction drastically increases the reliability of theoretical predictions for
such a class of systems as $^{16}$Ne/$^{16}$C on the proton side of the isobar.

%===============================================================================

\subsection{Widths of the $^{16}$Ne $0^+$ and $2^+$ states}

%===============================================================================

\begin{figure}
\begin{center}
\includegraphics[width=0.252\textwidth]{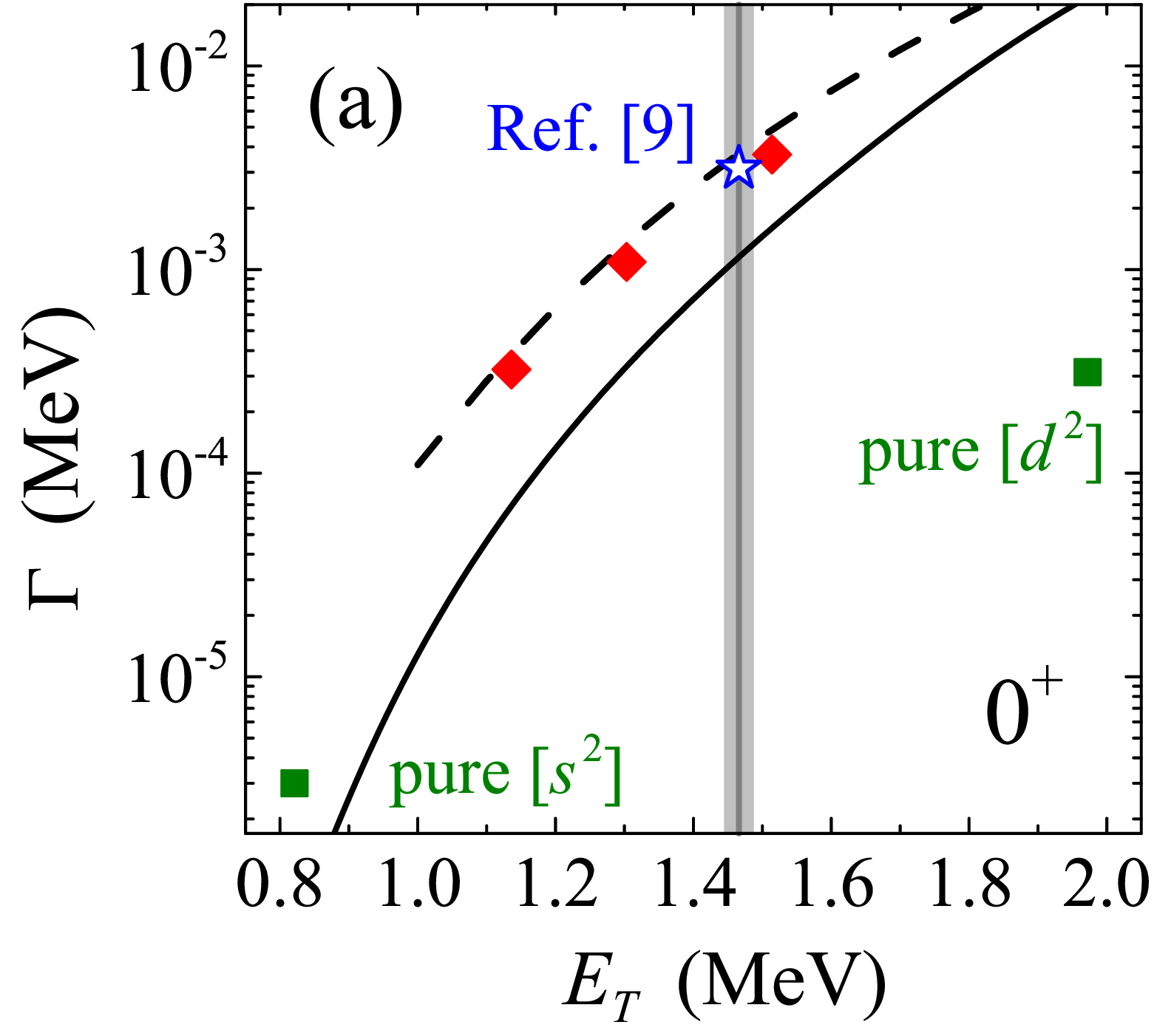}
\includegraphics[width=0.225\textwidth]{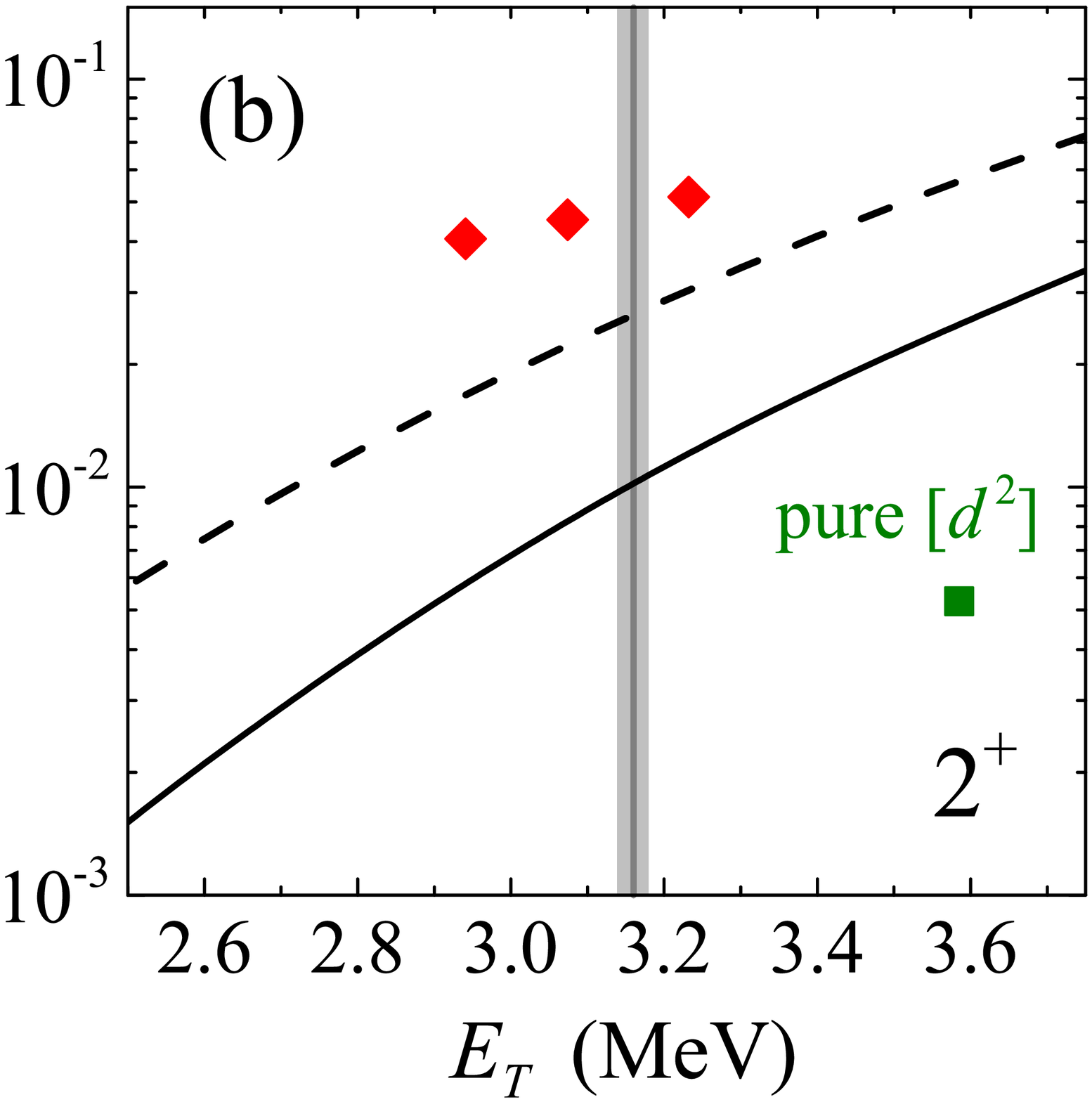}
\end{center}
\caption{The widths of the $^{16}$Ne $0^+$ and $2^+$ states. Solid and dashed
curves are three-body calculations and diproton model estimates from Ref.\
\cite{Grigorenko:2002}. Red diamonds are results of calculations with different
$E_r$ values from Table \ref{tab:poten}. Green squares shows the results of
calculations with limiting
cases of nuclear structure: pure $[s^2]$ or pure $[d^2]$. Blue star shows the
$0^+$ result from Ref.\ \cite{Brown:2014}. Vertical gray lines and dashed areas
correspond to experimental energies Ref.\ \cite{Brown:2014} and their
uncertainties.}
\label{fig:widths}
\end{figure}

The results of width calculations, see Table \ref{tab:poten}, are visualized in 
Figure \ref{fig:widths}. They are accompanied with the calculational results 
from Ref.\ \cite{Grigorenko:2002}. The latter work was one of the first of our 
works on the topic and the results provided there suffered from some technical 
issues, which were later overcome \cite{Pfutzner:2012}.

It can be seen that our new results for the $0^+$ state (red diamonds in Fig.\ 
\ref{fig:widths}a) are around a factor four larger than those from Ref.\ 
\cite{Grigorenko:2002} (solid curves in Fig.\ \ref{fig:widths}). However, they 
follow the trend provided by the old prediction very well. So, the difference is 
a pure convergence issue, connected to the small-basis calculations a decade 
ago.

A different situation is found for the $2^+$ state widths. The new calculational
results differ from the old ones even more and here they clearly follow a
different energy trend. Here we have to conclude that the three-body width
increase  connected with decrease of the $1/2^+$ state energy in the $^{15}$F
subsystem ``outweights'' the three-body width decrease  connected with
corresponding shift to lower $E_T$ energies.

The width results for the pure $[d^2]$ configuration (green squares in Fig.\ 
\ref{fig:widths}) are always more than an order of the magnitude lower than 
value expected by continuing the trend for realistic structure calculations or 
calculations with $[s^2]$ dominance. This is an indication of the uncertainty 
range for the two-proton decay widths, which, in principle, can be associated 
with unknown nuclear structure.

%-------------------------------------------------------------------------------
%-------------------------------------------------------------------------------

%===============================================================================

\section{Experimental data on $^{16}$Ne $0^+$ and $2^+$ states}

%===============================================================================

Figure \ref{fig:tes-ot-er} demonstrated the consistency of our theoretical TES
values to the most recent experimental data \cite{Brown:2014} on $^{16}$Ne
$0^+$ and $2^+$ states. The situation is, however, different for the other data.

The available experimental data on $^{16}$Ne $0^+$ and $2^+$ are listed in Table 
\ref{tab:exp}. The first thing which should be noted is that already the ground 
state data are still quite uncertain, spanning from 1.33 to 1.47 MeV (we omit 
here one of the early and imprecise results). This uncertainty is often larger 
than the provided errors of particular experiments. Even the most recent 
experiments \cite{Wamers:2014,Brown:2014}, both declaring to have the best 
precisions ever, disagree with each other for $E_T$ values of $0^+$ state beyond 
the provided errors. Thus the overall experimental situation is unsatisfactory 
already for the $0^+$ $^{16}$Ne ground state energies.

\begin{figure}
\begin{center}
\includegraphics[width=0.252\textwidth]{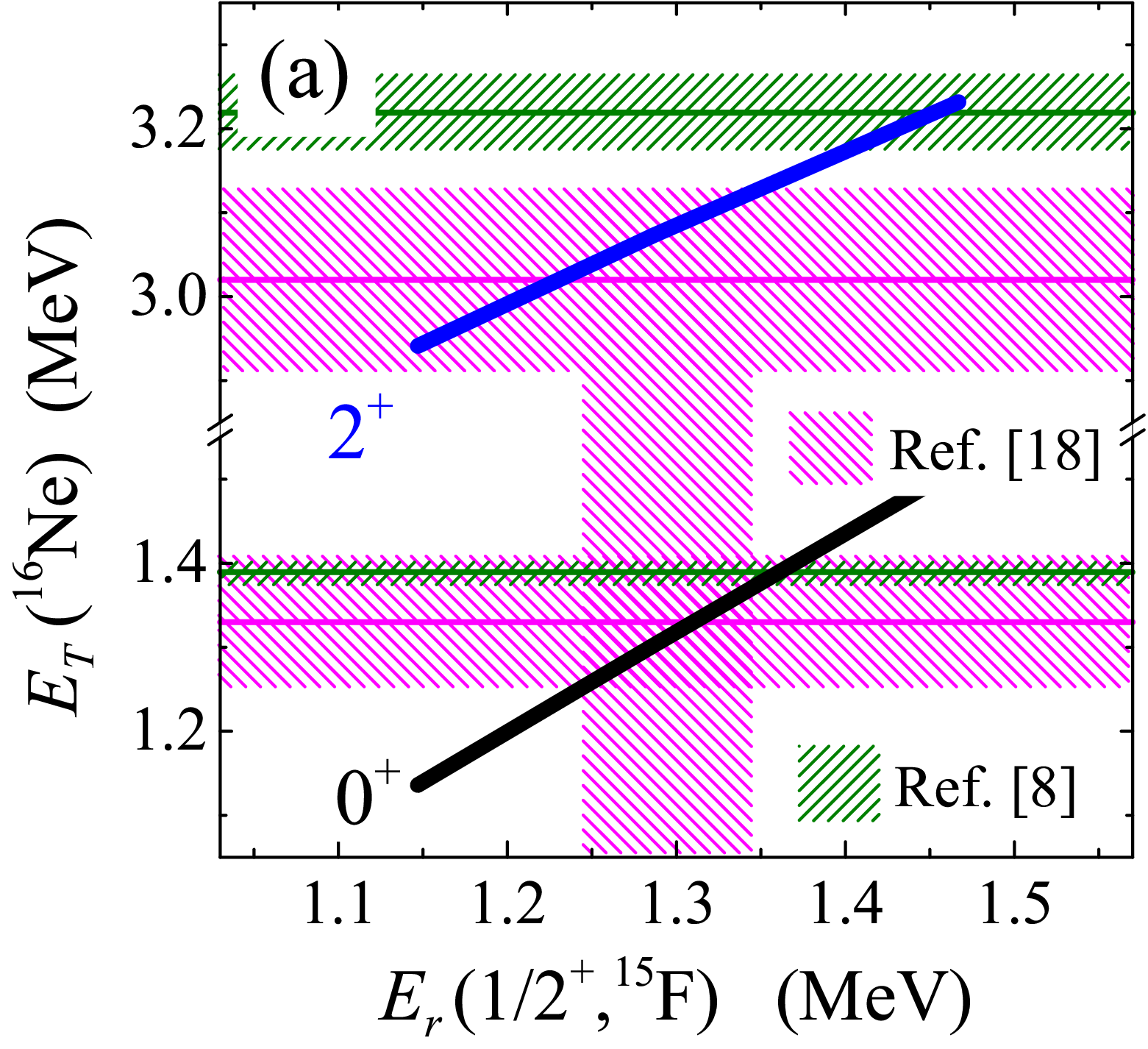}
\includegraphics[width=0.225\textwidth]{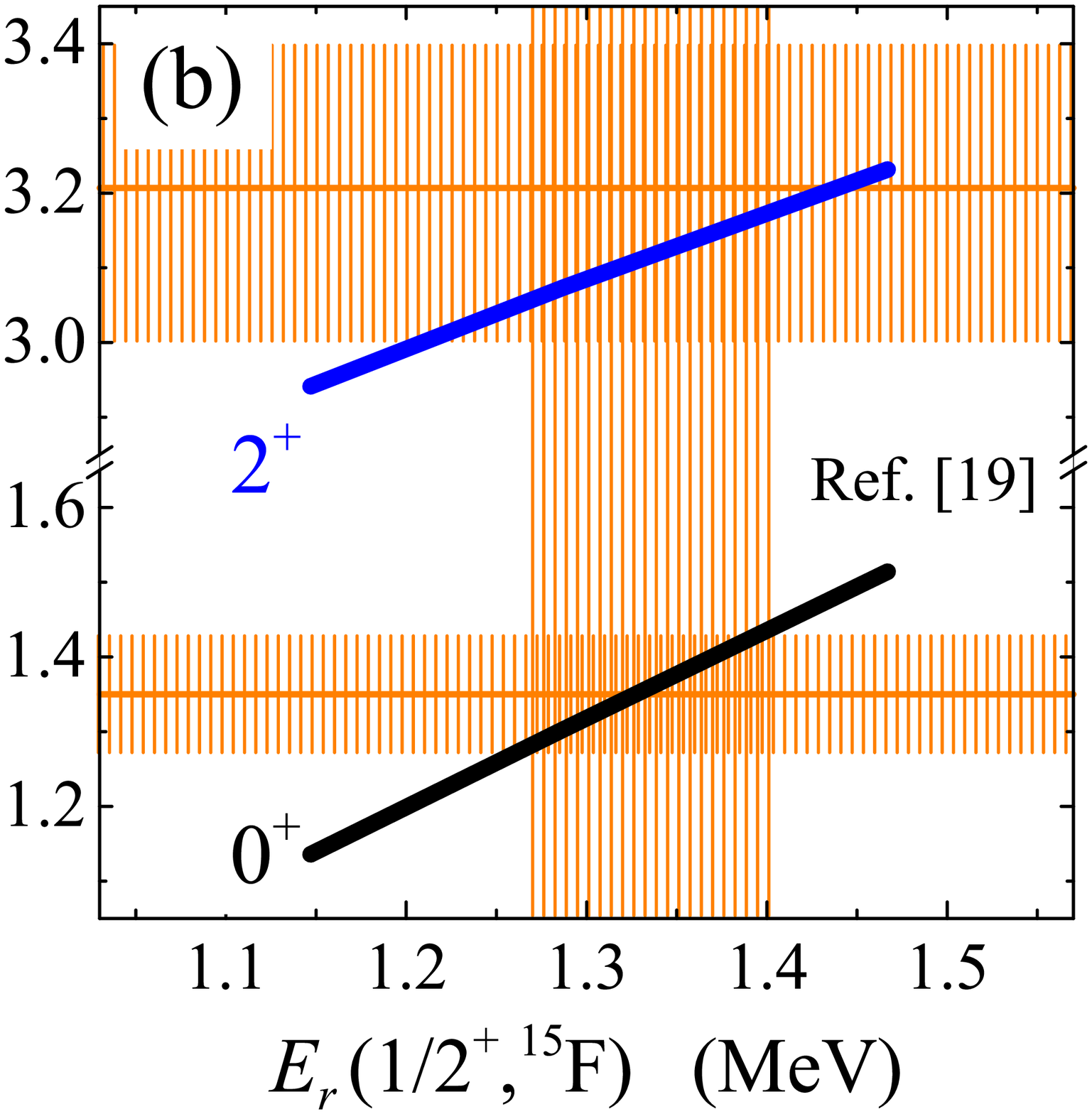}
\end{center}
\caption{Dependence of the $0^+$  and $2^+$ state energies in $^{16}$Ne on
the position of the g.s.\ $1/2^+$ resonance in $^{15}$F. The solid lines are the
same as in Fig.\ \ref{fig:tes-ot-er}. Horizontal gray lines
and hatched areas show the experimental values from Refs.\
\cite{KeKelis:1978,Wamers:2014} (a) and Ref.\ \cite{Mukha:2009} (b) with their
uncertainties. The vertical hatched areas indicate the $E_r$ ranges, where a
consistency between theory and experiment is achieved.}
\label{fig:consist-exp}
\end{figure}

Among the experimental data given in Table \ref{tab:exp}, the four experiments 
enlisted first \cite{KeKelis:1978,Mukha:2009,Wamers:2014,Brown:2014} provide 
both the $0^+$ and $2^+$ positions and thus allow to consider a consistency of 
these data with theoretical TES results as was done for \cite{Brown:2014} in 
Figure \ref{fig:tes-ot-er}. Such a comparison is provided in Fig.\ 
\ref{fig:consist-exp}. The consistency range for experiment \cite{Mukha:2009} 
$E_r=1.27-1.4$ MeV is quite broad and somehow overlaps with that found for the 
data of Ref.\ \cite{Brown:2014}. The consistency in the terms of TES exists for 
the data \cite{KeKelis:1978}. However, the obtained range of $E_r=1.25-1.35$ MeV 
is not compatible with that of \cite{Brown:2014}. Finally, the results of 
\cite{Wamers:2014} are not compatible in the TES terms as calculated in our work 
(the theoretical curves are crossed by the experimental ranges at somewhat 
different $E_r$ ranges). It should be noted that if a larger uncertainty is 
assumed for the g.s.\ energy in this experiment, than the consistency with 
theoretical results would be achieved at $E_r \sim 1.39$ MeV, also in agreement 
with \cite{Brown:2014}.

%===============================================================================
\begin{table}[b]
\caption{Experiments in which the properties of $0^+$ g.s.\ and first $2^+$
states in $^{16}$Ne were measured. Energies and widths are in MeV. The last
column provides the range of the $^{15}$F g.s.\ energies $E_r(1/2^+)$ in which 
the consistency of theoretical TES values can be achieved for $0^+$ and $2^+$ 
states simultaneously, see Figs.\ \ref{fig:tes-ot-er} and 
\ref{fig:consist-exp}.}
\begin{ruledtabular}
\begin{tabular}[c]{cccccc}
Ref.\ & $E_T(0^+)$ & $\Gamma(0^+)$ & $E_T(2^+)$ & $\Gamma(2^+)$ & $E_r(1/2^+)$
\\
\hline
\cite{KeKelis:1978} & 1.33(8) & 0.2(1) & 3.02(11) &        & $1.25-1.35$ \\
\cite{Mukha:2009}   & 1.35(8) &        & 3.2(2)   & 0.2(2) & $1.27-1.4$  \\
\cite{Wamers:2014}  & 1.388(15) & 0.082(15) & 3.220(46) & $< 0.05$  & none  \\
\cite{Brown:2014}   & 1.466(20) & $< 0.08$  & 3.16(2) & 0.02(1) & $1.39-1.42$
\\
\hline
\cite{Holt:1977}    & 1.8(5)    &    &    &    &    \\
\cite{Burleson:1980}& 1.466(45) &    &    &    &    \\
\cite{Woodward:1983}& 1.399(24) & 0.11(4) &    &    &    \\
\end{tabular}
\end{ruledtabular}
\label{tab:exp}
\end{table}
%===============================================================================

We have demonstrated above in Figs.\ \ref{fig:dep-et-ot-ws2} and 
\ref{fig:consist} that increase in precision of experimental data on state 
positions is required to make a better use of the TES results even for the most 
recent data of Ref.\ \cite{Brown:2014}. The Fig.\ \ref{fig:consist-exp} shows 
that the overall situation is even worse and we see that from experimental side 
also a broad controversy concerning the older data which should be resolved in 
general before definitive conclusions on actual TES behavior would become 
possible.

%-------------------------------------------------------------------------------
%-------------------------------------------------------------------------------

%===============================================================================

\section{Theoretical discussion}

%===============================================================================

The obvious way to use TES for nuclear structure studies in the $sd$-shell
systems with even number of valence nucleons is to apply it to derivation of
the configuration mixing rates. The basic idea was discussed in the Introduction
related to Eq.\ (\ref{eq:sd-wf}). This way of reasoning about the nuclear
structure was elaborated in a number of papers for various mirror pares of
$sd$-shell systems: $^{12}$O/$^{12}$Be \cite{Sherr:1999}, $^{15}$Ne/$^{15}$B
\cite{Fortune:2013}, $^{16}$Ne/$^{16}$C \cite{Ogawa:1999,Fortune:2002},
$^{17}$Ne/$^{17}$N \cite{Fortune:2001}, $^{18}$Ne/$^{18}$O
\cite{Sherr:1998,Ogawa:1999}.

It is necessary to note that the connection between TES and the configuration 
mixing is straightforward and simple only in the case of an independent particle 
model with well defined orbital characteristics. However, even in the 
independent particle model fixing of orbital sizes requires precise knowledge of 
excitation energies of the single particle states for $A-1$ ``subsystem'' on the 
proton side of the isobar. Among the mentioned systems this is not the case for 
$^{12}$O, $^{15}$Ne, and $^{16}$Ne, with ``subsystems'' $^{11}$N, $^{14}$F, 
$^{15}$F which possess quite broad $s_{1/2}$ states making the precise 
experimental determination of CDE problematic. In this work we demonstrate by 
the example of $^{16}$Ne that this issue have a major impact on conclusions 
about the configuration mixing, see Fig.\ \ref{fig:consist}. On top of this 
issue we also insist on the existence of the dynamic three-body mechanism of TES 
leading to a strong modification of the configuration mixing rates when we move 
from neutron to proton side of the isobar.

Paper \cite{Fortune:2013} provides an impressive prediction of the $^{15}$Ne 
g.s.\ energy $E_T=2.68(24)$ MeV which appears to be in a good agreement with the 
later measured value $E_T=2.522(66)$  MeV \cite{Wamers:2014}. This prediction is 
based on two ingredients. (i) Phenomenological linear dependence on $W(s^2)$ for 
``scaled CDE'' $(S_{2n}-S_{2p})A^{1/3}/Z_<$ was derived in \cite{Fortune:2013} 
based on the data for several $Z=8,\, 10$ isobaric mirror partner pairs. (ii) A 
plausible value $W(s^2)=66 \%$ for $^{15}$B was assumed in \cite{Fortune:2013} 
just between the $W(s^2)=86 \%$ deduced for $^{14}$Be and $W(s^2)=46 \%$ for 
$^{16}$C. We point to the need to reconcile this type of phenomenology with more 
complicated dependencies obtained in this work. The dependence (i) is in 
principle analogous to the dependence of Fig.\ \ref{fig:dep-et-ot-ws2}(a). 
However, we obtain a set of such dependencies even for one single nucleus 
$^{16}$Ne depending on $E_r$ in $^{15}$F. Studies of the configuration mixing in 
$^{15}$Ne in a three-body model could also elucidate the issue (ii).

%\vspace*{5mm}

%===============================================================================

\section{Conclusions}

%===============================================================================

The following  main results are obtained in this work.

\noindent (i) Large isospin symmetry breaking on the level of the nuclear
structure associated  with TES was predicted in Ref.\ \cite{Grigorenko:2002} for
$0^+$ states and further elaborated in this work also in the case of $2^+$
states. We have found that in the $^{16}$Ne/$^{16}$C mirror pair the ``dynamic''
component of TES, connected to the structure modification, is responsible for
about a half of the whole TES effect. The scale of the structure modification in
these  mirror nuclei is $20-25 \%$ for the $0^+$ ground states and $6-10 \%$ for
the first $2^+$ states.

\noindent (ii) In this work we study carefully the stability of such predictions
to theoretical inputs to the calculations. In our predictions the structure of
the $^{16}$Ne states appears to be very stable to the admissible variation of
parameters. Quite unexpectedly, the stability of predictions for $^{16}$Ne is
much better than for $^{16}$C, presumably due to a more peripheral character of
its WF.

\noindent (iii) Accurate studies of the Coulomb displacement energies indicate
that a consistency among three parameters should be requested: the decay
energy $E_T$, the $^{15}$F g.s.\ energy $E_r$, and the configuration mixing
parameters [$W(s^2)/W(d^2)$ for $0^+$ and $W(sd)/W(d^2)$ for $2^+$ states]. This
is a more complicated dependence than it is ordinarily assumed. Typically
TES is correlated with the configuration mixing \emph{only}, to provide
predictions about nuclear structure
\cite{Sherr:1998,Sherr:1999,Ogawa:1999,Fortune:2001,Fortune:2002,Fortune:2013}.

\noindent (iv) The energy of the $^{15}$F $1/2^+$ g.s.\ extracted from our
analysis is $E_r=1.39-1.42$ MeV. Some shift to the lower energies is possible
due to WF configurations which are beyond our model. The $E_r$ values above
$1.43-1.45$ MeV are practically excluded by our analysis.

\noindent (v) The current experimental situation is too uncertain for 
high-precision comparison with the calculated results. Even the very accurate 
values of the $^{16}$Ne experimental energies of Ref.\ \cite{Brown:2014} allow a 
consistency with theoretical predictions of TES in a broad range of other 
parameters. Further increase in the precision of the measured energies in 
$^{15}$F and $^{16}$Ne would impose very stringent limits in the parameter space 
in which the consistency with theory is possible. Thus TES could become a 
sensitive tool for extraction of deep structural information about 
$^{16}$Ne/$^{16}$C mirror nuclei as well as $sd$-shell nuclei with analogous 
dynamics.

%===============================================================================
%
\textit{Acknowledgments.}
%
%===============================================================================
%
--- L.V.G.\ is supported by the Russian Foundation for Basic Research 
14-02-00090-a and Ministry of Education and Science NSh-932.2014.2 grants. We 
are grateful to Prof.\ J.S.\ Vaagen for careful reading of the manuscript and 
useful comments.

%###############################################################################

\bibliographystyle{apsrev}
\bibliography{c:/latex/all}

%###############################################################################

\end{document}